\documentstyle[psfig,epsf]{article}
\begin{document}
\def\ap#1#2#3   {{\em Ann. Phys. (NY)} {\bf#1} (#2) #3}   
\def\apj#1#2#3  {{\em Astrophys. J.} {\bf#1} (#2) #3.} 
\def\apjl#1#2#3 {{\em Astrophys. J. Lett.} {\bf#1} (#2) #3.}
\def\app#1#2#3  {{\em Acta. Phys. Pol.} {\bf#1} (#2) #3.}
\def\ar#1#2#3   {{\em Ann. Rev. Nucl. Part. Sci.} {\bf#1} (#2) #3.}
\def\cpc#1#2#3  {{\em Computer Phys. Comm.} {\bf#1} (#2) #3.}
\def\epj#1#2#3  {{\em Europ. Phys. J.} {\bf#1} (#2) #3}
\def\err#1#2#3  {{\it Erratum} {\bf#1} (#2) #3.}
\def\ib#1#2#3   {{\it ibid.} {\bf#1} (#2) #3.}
\def\jmp#1#2#3  {{\em J. Math. Phys.} {\bf#1} (#2) #3.}
\def\ijmp#1#2#3 {{\em Int. J. Mod. Phys.} {\bf#1} (#2) #3}
\def\jetp#1#2#3 {{\em JETP Lett.} {\bf#1} (#2) #3}
\def\jpg#1#2#3  {{\em J. Phys. G.} {\bf#1} (#2) #3.}
\def\mpl#1#2#3  {{\em Mod. Phys. Lett.} {\bf#1} (#2) #3.}
\def\nat#1#2#3  {{\em Nature (London)} {\bf#1} (#2) #3.}
\def\nc#1#2#3   {{\em Nuovo Cim.} {\bf#1} (#2) #3.}
\def\nim#1#2#3  {{\em Nucl. Instr. Meth.} {\bf#1} (#2) #3.}
\def\np#1#2#3   {{\em Nucl. Phys.} {\bf#1} (#2) #3}
\def\pcps#1#2#3 {{\em Proc. Cam. Phil. Soc.} {\bf#1} (#2) #3.}
\def\pl#1#2#3   {{\em Phys. Lett.} {\bf#1} (#2) #3}
\def\prep#1#2#3 {{\em Phys. Rep.} {\bf#1} (#2) #3}
\def\prev#1#2#3 {{\em Phys. Rev.} {\bf#1} (#2) #3}
\def\prl#1#2#3  {{\em Phys. Rev. Lett.} {\bf#1} (#2) #3}
\def\prs#1#2#3  {{\em Proc. Roy. Soc.} {\bf#1} (#2) #3.}
\def\ptp#1#2#3  {{\em Prog. Th. Phys.} {\bf#1} (#2) #3.}
\def\ps#1#2#3   {{\em Physica Scripta} {\bf#1} (#2) #3.}
\def\rmp#1#2#3  {{\em Rev. Mod. Phys.} {\bf#1} (#2) #3}
\def\rpp#1#2#3  {{\em Rep. Prog. Phys.} {\bf#1} (#2) #3.}
\def\sjnp#1#2#3 {{\em Sov. J. Nucl. Phys.} {\bf#1} (#2) #3}
\def\shep#1#2#3 {{\em Surveys in High Energy Phys.} {\bf#1} (#2) #3}
\def\spj#1#2#3  {{\em Sov. Phys. JEPT} {\bf#1} (#2) #3}
\def\spu#1#2#3  {{\em Sov. Phys.-Usp.} {\bf#1} (#2) #3.}
\def\zp#1#2#3   {{\em Zeit. Phys.} {\bf#1} (#2) #3}

\hyphenation{author another created financial paper re-commend-ed}
\begin{center}
{\Large \bf PHOTON STRUCTURE} \\

\vspace{4mm}

Jacek Turnau for the H1 and ZEUS Collaborations\footnote{To be published in
Proceedings of the Workshop on Quantum Field Theory and High Energy Physics
Tver 2000, Work partially supported
by Polish Committee for Scientific Research grant 2P03B10318.}\\
H. Niewodniczanski Institute of Nuclear Physics, Krakow\\
Kawiory 26a 30-055 Krakow e-mail jacek.turnau@ifj.edu.pl\\
\end{center}

\begin{abstract}
Measurements of  di-jet production at HERA provide an important test of 
NLO QCD calculations. It is also a source of information about the
partonic content
of the photon, complementary to the measurements in $e^+e^-$ experiments. In
this article we review the status of the photon structure studies with 
particular emphasis on HERA measurements.
\end{abstract}
\section{Introduction}
The HERA collider, where 27.5 GeV positrons collide with 820/920 GeV protons, 
is traditionally considered as a natural extension of Rutherford's 
experiment and the process of deep inelastic scattering ($ep$ DIS) 
is interpreted as a reaction
 in which a virtual photon, radiated by the incoming positron, probes the 
structure of the proton. However, there are regions of phase space for $ep$ 
collisions where it is other way round: where partons in the proton probe 
the photon structure.   
\par
The structure of the talk will be the following: it will start with
posing the problem i.e. a description of our  understanding of the 
structure of the photon in terms of Quantum Field Theory (QFT)  and the 
motivations for its investigation. Next, we briefly review measurements 
of the photon 
structure in $e^+e^-$ experiments and its general properties. In the  
main part,  we review HERA measurements sensitive to the  structure of the 
photon. 
We end with a summary and conclusions.
\par
Various aspects of the subject discussed here have been covered
recently in many excellent review articles e.g. \cite{dainton},
\cite{krawczyk},
\cite{nissius}, \cite{alevi}.
\section{Structure or interaction dynamics?}
The photon is the gauge boson of QED and, as far as we know, elementary.
In any  QFT the existence of the interaction means also that the quanta 
themselves
have structure. Coupling between field quanta will make possible fluctuations,
or splitting of any quantum over limited time, into two (or more in higher 
order) quanta. For example in QED a photon can fluctuate into an 
electron-positron ($e^+e^-$) pair, or, as first noted by 
Ioffe \cite{ioffe}, into a $q\bar{q}$ pair.
If the fluctuation time defined in the probing objects rest frame as
$t_f \approx 2E_{\gamma}/m_{q\bar{q}}^2$ is much larger than the interaction 
time $t_{int}$, the photon builds up structure in the interaction. Here, 
$E_{\gamma}$ is the energy of the fluctuating photon and $m_{q\bar{q}}$ is the
 mass into which it fluctuates. The fluctuation time of a photon with 
virtuality $Q^2$ is given by $t_f \approx 2E_{\gamma}/(m_{q\bar{q}}^2+Q^2)$,
and thus at very high $Q^2$ one does not expect the condition 
$t_f \gg t_{int}$, unless the probe has even higher virtuality. 
\par
It is clear that what we described 
here as the photon structure is a part of the interaction dynamics. In practice
in any high energy interaction, the distinction between structure and 
interaction is determined by the factorisation scale. This is a momentum 
transfer
squared scale below which any parton activity is considered to be part of the 
parent structure and above which any parton activity is considered to be a part
of the interaction dynamics. It is exactly in this sense that in  $ep$ DIS 
we measure the proton structure function
$F_2(x,Q^2)$ at a factorisation (or resolution) scale $Q^2$: $Q^2$ is almost
always the largest momentum transfer squared scale in DIS, so all parton 
activity at lower transverse momenta should be associated with the target i.e.
proton. The structure function is interpreted as a density of quarks and 
antiquarks which carry a fraction $x$ of the proton momentum. Any structure 
function is thus dependent on the choice of the factorisation (resolution) 
scale. 
It was shown that structure functions for hadrons can be used universally
in any high energy interaction at a given scale. In the same way it is also
expected that photon structure $F_2^{\gamma}(x_{\gamma},Q^2)$ will be similarly
universal and factorisable.
\par
When interactions are switched off the photon remains structureless so we say 
that photon structure is driven by fluctuations. It is in many respects 
different from that of hadrons which have structure even after interactions 
are switched off (valence driven structure). In addition, the photon 
structure can be investigated as a function of its virtuality i.e. size, which 
is not possible for hadrons. We have thus several strong motivations for the 
photon structure studies:
\begin{itemize}
\item
Measurements of the photon structure function provide interesting tests of QCD.
Some pose even a challenge to theory e.g. there is no full theory of the 
parton density suppression due to increasing virtuality (only an asymptotic 
prediction exists)
\item
The parametrisation of the universal and factorisable photon structure
function based on LEP and HERA measurements is a necessary component of the 
description of many high energy processes
\item
At future linear colliders it will be very important to understand the
large number of events from photon interactions. Some photon interactions
can be important probes to look for new physics, for example the production
of Higgs bosons at a photon linear collider ($\gamma\gamma \rightarrow H$)
\cite{melles}.
\end{itemize}
\section{Photon structure from $e^+e^-$ and its general properties}
The hadronic structure of the photon can be probed either by virtual photons in
$e^+e^-$ collisions or by virtual partons in $ep$ collisions at HERA. This 
article
is devoted mainly to the recent measurements at the HERA collider, but for the
sake of completeness some $e^+e^-$ results are  discussed in this section.
\par
\begin{figure}[htb]
\vspace{-0.5cm}
\begin{center}
\psfig{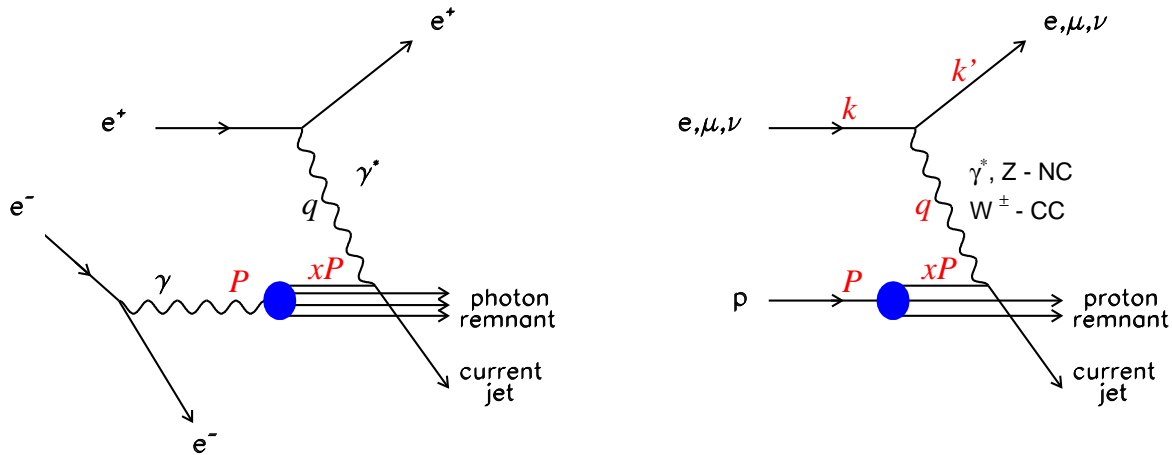}
\end{center}
\vspace*{-0.5cm}
\caption{A diagram  (left) describing the process of deep inelastic 
scattering of a positron  on a quasi-real $\gamma$ using the reaction 
$e^+e^- \rightarrow e^+e^-X$.
The four-vector of the virtual $\gamma$ (probe) is $q$, the four momentum 
of the 
struck quark is $xP$. To underline the analogy with the measurement of the
hadronic structure function, a diagram describing lepton-proton deep 
inelastic scattering is shown in the right diagram.}
\label{fig:photon-dis}
\end{figure}

The principle of the measurement of the hadronic structure function of the 
photon
$F_2^{\gamma}(x_{\gamma},Q^2)$ is depicted in Figure \ref{fig:photon-dis}
 (left). 
A highly virtual $\gamma^{\star}$ with large $Q^2=-q^2$ probes a quasi-real
$\gamma$ with virtuality $P^2 \approx 0$. In this measurement the energy of the 
target is not known so $x_{\gamma}$ (the fraction of the photon momentum 
carried by the interacting parton) has to be reconstructed from the hadronic 
final 
state. This leads to systematic uncertainties and a limited reach for small
values of $x_{\gamma}$. To underline the analogies and differences we depict
the principle of the measurement of the proton structure function on the 
right of 
Figure \ref{fig:photon-dis}.
\par
Many measurements of the hadronic structure function of the photon have been 
performed at several $e^+e^-$ colliders. The range of resolution scale covered 
by various experiments is $0.24 \leq \langle Q^2 \rangle \leq 700 ~{\rm GeV}^2$.
\begin{figure}[htb]
\begin{center}
\psfig{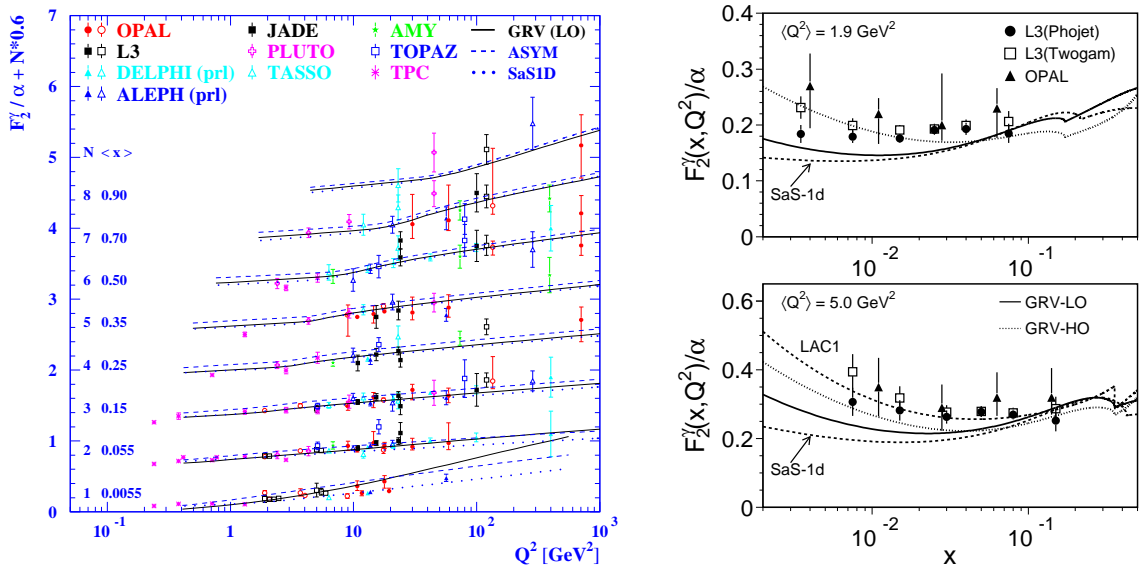}
\end{center}
\vspace*{-0.5cm}
\caption{{\em Left:}
The photon structure function $F_2^\gamma$, as a function of $Q^2$, for
average $x$ values as given in the figure. The curves are the
expectations of different parametrisations of parton distributions in
the photon. {\em Right:} The photon structure as a function of the photon
momentum fraction $x$ at fixed resolution scale $Q^2$. The curves are 
different parametrisations of parton distributions in the photon.}
\label{fig:eef2}
\end{figure}
In Figure \ref{fig:eef2}  we present a compilation \cite{nissius} (recently
updated with OPAL data \cite{OPAL2}) of photon 
structure function measurements in various $e^+e^-$ experiments. $F_2^{\gamma}$
is plotted as a function of $Q^2$ for several ranges of $x_{\gamma}$. One can
see positive scaling violation for all $x_{\gamma}$. This is the most striking
feature of the photon structure in comparison to the structure of the proton,
for which positive scaling violation is observed only at low $x$ values.
This different behavior can be understood in terms of perturbative QCD as 
coming from the splitting $\gamma \rightarrow q\bar{q}$, having no analogue 
in the proton case. In addition, and again contrary to the proton case,
it also causes the photon structure function to be large for high $x$ values.
\par
A simple LO parametrisation of the photon structure function
\begin{equation}
F_2^{\gamma} =3 \sum_{q}e_q^4\frac{\alpha}{\pi}x[x^2+(1-x)^2]{\rm ln}\frac{Q^2}
{\Lambda^2} + [{\rm'hadron'} \equiv {\rm VDM}]
\label{eq:witten1}
\end{equation}
can be written down in QCD. The expression is asymptotic i.e. all terms
not multiplied by large ${\rm ln}Q^2$ are dropped. The second term in the equation 
 -- ``hadron'' part, corresponds to $\gamma \rightarrow q\bar{q}$ fluctuations
with virtuality below a cutoff value $\Lambda^2$. This noncalculable piece
is traditionally represented by photon fluctuations into vector mesons, as
proposed 30 years ago in terms of the Vector Dominance Model \cite{VDM}. 
Witten
\cite{witten} and others showed that the above mentioned features of the LO 
formula
 (1) remain valid in calculations which include higher order perturbative
terms. Thus a rising dependence of $F_2^{\gamma}$ with increasing probe scale
$Q^2$ throughout most of the range of $x$ is thereby seen to be a 
characteristic
feature of the ``fluctuation driven'' (rather than ``valence driven'') 
structure functions. 
Measurements compiled in Figure \ref{fig:eef2} provide therefore an 
interesting and important test of pQCD.
\par
Since the measurements of $F_2^{\gamma}$ span a wide range of $Q^2$, it is  
possible to apply the DGLAP pQCD formalism \cite{altar} as is 
done for the  proton structure function. As a result parton density
functions (pdfs) for the photon are available.
\par
There exist several pdfs for real and also for virtual photons in leading
and next-to-leading order, which are based on the full evolution equations. 
They
are constructed in manner very similar to that for  the proton pdfs. They 
differ in the assumptions 
made about the starting scale, the input distributions assumed at this scale
 and also in the amount of data used in fitting their parameters. Those most 
quoted in the literature are LAC \cite{LAC}, GRV \cite{GRV}, AGF 
\cite{AGF} and SAS \cite{SAS}.
These parametrisations were developed before the publication of the new 
LEP data.
LEP has measured the photon structure function $F_2^{\gamma}$ in the range
$0.002 \leq x_{\gamma} \leq 1.0 $ and 
$1.86 < \langle Q^2 \rangle < 700 ~{\rm GeV}^2$. These
data represent an important step in the reduction of statistical and systematic
uncertainties and cover a larger kinematic region both in $x_{\gamma}$ 
and $Q^2$.
From Figure \ref{fig:eef2}  (right) it is clear that the LEP data provide  new
information on photon structure and the existing pdf parametrisations will 
have to be revisited. 
\section{Photon structure from HERA}
At HERA the photon structure is investigated mainly by measurements of high
$E_T$ jets and charged particles \cite{H11}-\cite{Z3}. The partonic 
structure of quasi-real or virtual target photons is probed by a parton from 
the proton producing a pair of partons of large transverse momentum squared
$E_T^2 $ much larger than the photon virtuality $Q^2$. Due to the large cross 
section at HERA
the photon can be probed at even larger factorisation scales than at LEP.
\par
The photoproduction of jets through a hard parton differs from the jet 
production in hadron-hadron collisions in one respect: the photon interacts
either by means of its direct coupling to high $E_T$ partons or by means
of partons in its structure, which are resolved in the interaction.
\begin{figure}[htb]
\begin{center}
\centerline{\psfig{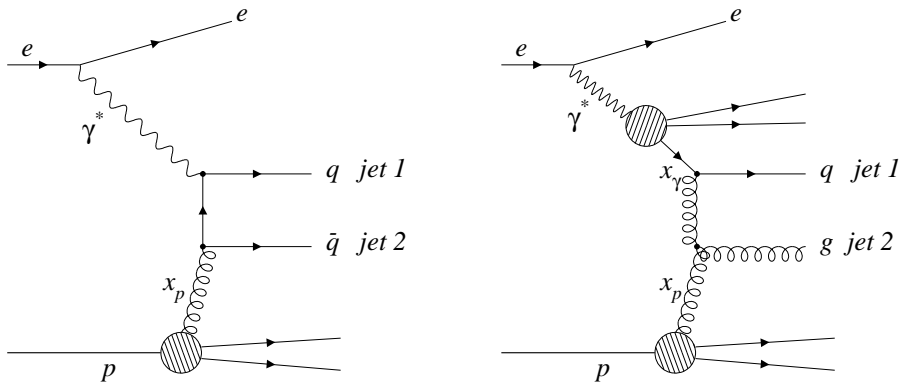}}
\end{center}
\vspace*{-0.5cm}
\caption{A diagram  (left) describing the direct process of di-jet production
in LO. The photon couples directly to high $E_T$ partons. The diagram
on the right depicts the resolved di-jet production in LO. The photon interacts
through partons belonging to its structure which carry $x_{\gamma}$ fraction
of its momentum.}
\label{fig:dir-res}
\end{figure}
This is depicted in Figure~\ref{fig:dir-res}. The terminology ``direct'' 
and ``resolved''
processes is now established, though in reality the direct photon contains 
fluctuation driven
structure not resolved at a given scale.
\par
The fraction of the photon momentum carried by an interacting parton can be 
reconstructed at
hadron level from jet kinematics
\begin{equation}
x_{\gamma}^{jet} = \sum_{jets}\frac{E_{Tjet} e^{-\eta_{jet}}}{E_{\gamma}}  
\label{eq:xgamma}
\end{equation}
where $E_{Tjet}$ and $\eta_{jet}$ are the transverse jet energy and 
pseudorapidity  of the jets. The 
sum extends over high $E_T$ jets (i.e. does not include the photon and proton 
remnant jets).
The ``true'' (i.e. parton level) fraction $x_{\gamma}$ can be unfolded 
from the data using
simulated $x_{\gamma} - x_{\gamma}^{jet}$ correlations. In order to reach low values of
$x_{\gamma}$ the measurement should extend to low values of the jet transverse momentum and/or
high values of their pseudorapidity (i.e. close to the proton remnant). Both requirements
are experimentally challenging. In particular, the low cut on $E_T$ results in large corrections
for hadronisation and soft underlying events i.e. secondary interactions of the partons 
belonging to the photon and proton remnants.
\par
In the LO QCD approximation to  di-jet photoproduction, the $ep$ cross section is written
\begin{equation}
\frac{{\rm d}^5\sigma}{{\rm d}y{\rm d}x_{\gamma} 
                       {\rm d}x_p {\rm d}\cos\theta^*  {\rm
                       d}Q^2} =   
  \frac{1}{32 \pi s_{ep}} 
\frac{f_{\gamma/{\rm e}}(y,Q^2)}{y} 
\sum_{ij} 
\frac{f_{i/\gamma}(x_\gamma,E_T^2,Q^2)}{x_\gamma} 
\frac{f_{j/p}(x_{\rm p},E_T^2)}{x_p} 
|M_{ij}(\cos\theta^*)|^2
\label{eq:fullsig}
\end{equation}
where $f_{i/\gamma}$  and $f_{i/p}$ are the pdfs for each parton species 
$i$ in a photon  and a proton, respectively. They are evaluated at the 
factorisation scale
$E_T^2$. The $M_{ij}$ are QCD matrix elements for 
$2\rightarrow 2$ parton-parton
hard scattering processes. The quantity $s_{ep}$ is the square of the centre of mass
energy in the $ep$ collision, and $\theta^*$ is the polar angle of the outgoing
partons in the parton-parton centre of mass frame.
\par
The measurement of the di-jet cross section can be exploited to gain information about the
 parton
densities in the photon. There are two conceptually different approaches, which can be 
nicknamed ``extraction'' and ``universality test''. 
\par
We can extract the parton density in the photon treating the measured 
di-jet cross 
section, the parton density in the proton (well constrained by inclusive $ep$ DIS data) and
 the QCD matrix element as  input to formula ~(\ref{eq:fullsig}). 
The result of the measurement is
a fundamental, parton level quantity. This approach has one 
important
limitation: it is the LO procedure, because it employs the LO notion of the 
effective parton density and the LO Monte Carlo which relates parton to hadron 
levels. The H1 collaboration extracted the parton density in photon  in the region of the 
low $x_{\gamma}$ where gluon component of the parton density is of particular 
interest and poorly constrained by $e^+e^-$ measurements. 
\par
In the ``universality test'' we have different objective: to check if 
the parametrisation of parton density in photon constrained by $e^+e^-$ 
measurements is adequate for the 
description of the inclusive di-jet cross section within NLO QCD theory.
In this approach one of the existing  parametrisations of the parton density 
in photon \cite{LAC}-\cite{SAS} is used  
as  input information in the NLO calculation \cite{kramer} of  jet 
photoproduction cross section at the  parton level. In order to be
consistent, we have to avoid the application of the LO Monte Carlo for 
hadronic  and soft underlying event corrections and  thus restrict the 
analysis to the 
region where these corrections are expected to be small: high transverse jet 
energy hence  high $x_{\gamma}$ (see formula ~(\ref{eq:xgamma})). 
The calculated parton level cross sections are compared with the data 
at the hadron level, which means that the data are corrected for detector 
effects only.
\par
It should be underlined that neither approach is satisfactory, especially 
at the level of the measurement
precision recently reached by both HERA experiments. 
Tools for extraction of the NLO pdf
in photons from di-jet measurements, similar to those existing for the proton  pdf, are missing.
Obviously, there are  important technical differences  between electromagnetic 
and strong probes: in the case of $ep$ or $e\gamma$ DIS the resolution scale  
does not have 
to be  unfolded for hadronisation and other QCD effects, because it is known 
directly from the electron measurement. In addition, the 
strong probe  comes in two species (quark or gluon), and this certainly 
complicates the fitting procedure.
The LO solution to  this problem,  i.e. Single Effective Parton Density 
approach (see sec. 4.1), may be not applicable in NLO theory.
\subsection{Measurement of the Effective Parton Density of quasi-real photons}
At HERA we cannot distinguish experimentally which partons initiated the hard scattering
process in Figure \ref{fig:dir-res}, therefore formula ~(\ref{eq:fullsig})
 is not suitable
for the extraction of parton densities. To avoid this difficulty, the H1 Collaboration 
adopted the Single Effective Subprocess (SES) approximation \cite{comb}, 
which exploits the fact 
that the dominant contributions to the
cross section come from the parton-parton scattering matrix elements which have
similar shapes. Introducing the effective pdf 
in the photon 
\begin{equation}
f^{{\rm eff}}_{\gamma}(x_{\gamma},E_T^2)\equiv 
\sum_{n_f}(f_{q/\gamma}(x_{\gamma},E_T^2)
+ f_{\bar{q}/\gamma}(x_{\gamma},E_T^2)) 
+ \frac{9}{4}f_{g/\gamma}(x_{\gamma},E_T^2),  
\end{equation}
the effective proton pdf $f^{{\rm eff}}_p$ and the SES matrix element
$M_{{\rm SES}}$, we can express the resolved photon contribution to
the cross section by the product $f^{{\rm eff}}_{\gamma}f^{{\rm eff}}_p|M_{{\rm SES}}|^2$.
As  $f^{{\rm eff}}_p$ is well constrained by the data and $M_{{\rm SES}}$ is calculable
in perturbative QCD, $f^{{\rm eff}}_{\gamma}$ can be directly determined from the 
measurement 
of the di-parton cross section. 
\par
\begin{figure}[htb]
\vspace{-1cm}
\begin{center}
\psfig{file=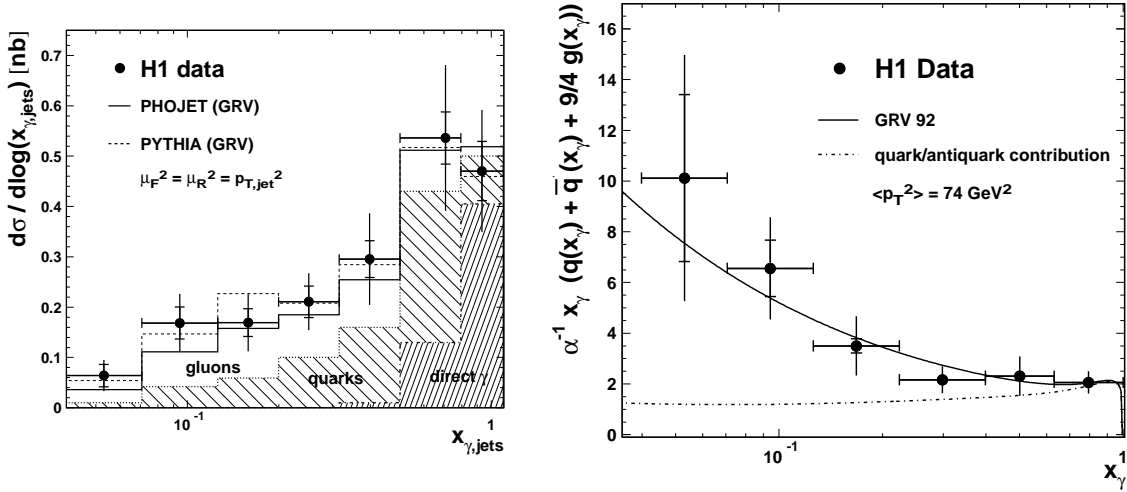,height=6.5cm}
\end{center}
\vspace*{-0.5cm}
\caption{{\em Left:} The di-jet cross section
as a function of the parton momentum fraction $x_{\gamma}^{jet}$. The histograms
represent a LO QCD calculation showing the contributions of the direct
photon-proton interactions as well as the quark and gluon induced processes 
using GRV pdfs for both photon and proton. {\em Right:} The $x_{\gamma}$ dependence of the effective parton
density of the real photon extracted from di-jet photoproduction; the curves superimposed
are the expectation from QCD LO GRV92 parametrisation of photon structure.}
\label{fig:q-real}
\end{figure}

In Figure \ref{fig:q-real}  (left) we show the   H1 measurement \cite{H13} 
of the di-jet cross section
as a function of the parton momentum fraction $x_{\gamma}^{jet}$.
 
The histograms
represent a LO QCD MC calculation showing the contributions of the direct
photon-proton interactions as well as the quark and gluon induced processes 
using the GRV 
pdfs for photon and proton. It should be noted that the relatively low cut
on the jet transverse energy ($E_T^{jets} >$ 6 ~{\rm GeV}) allows a precise 
measurement down to 
$x_{\gamma}=0.05$. Sensitivity to the gluon content of the photon is clearly
seen. In the definition of $d\sigma/dx_{\gamma}^{jet}$ pedestal of the
jet transverse momentum due to the soft underlying event is subtracted. 
This is carefully checked by comparing the energy flow around the jets in
data and simulation which agrees  very well. The remaining differences
are added to the systematic errors.
In Figure \ref{fig:q-real}  (right) we show the parton density in the photon 
extracted from the differential cross section. $x_{\gamma}^{jet}$ has been
unfolded to parton level $x_{\gamma}$ and the effective parton density is
determined by the comparison of MC and the data
\begin{equation}
f_{\gamma}^{DATA}=f_{\gamma}^{MC}\frac{\sigma^{DATA}}{\sigma^{MC}}.
\end{equation}

\begin{figure}[htb]
\vspace{-1cm}
\begin{center}
\centerline{\psfig{file=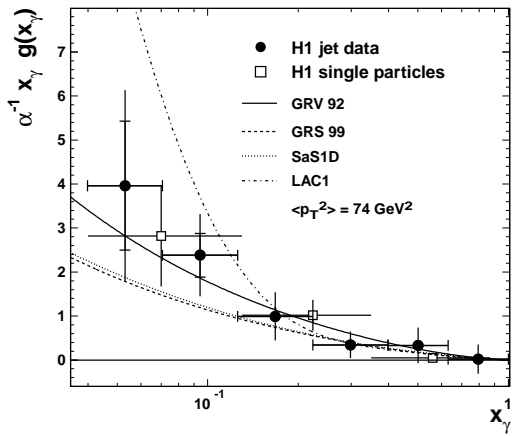,height=6.5cm}}
\end{center}
\vspace*{-1.1cm}
\caption{Gluon distribution $g(x_{\gamma})$ of the photon as a function
of the photon momentum fraction for a mean $\hat{p}_t^2=74 ~{\rm GeV}^2$ of hard
partons. The inner error bars give statistical error and outer error bars 
give the total error. The data points with open squares show a previous 
measurement \cite{H14} of H1 which used single high $E_T$ particles
to determine the LO gluon density of the photon at a mean 
$\hat{p}_t^2=38 ~{\rm GeV}^2$. The leading order parametrisations of the gluon 
distribution based on fits to $e^+e^-$ two-photon data are also shown.} 
\label{fig:gluon}
\end{figure}

Figure \ref{fig:gluon} shows the  gluon content of the photon obtained from 
the effective photon pdf by subtraction of the GRV quark contribution.
A good agreement with the gluon density derived from the photoproduction 
of high
transverse momentum charged particles \cite{H14} should be noted, as these
 two measurements have different systematics. Thus the large gluon component
of the photon pdf at low values of $x_{\gamma}$ may be considered as 
established. 
The systematic errors dominate
above $x_{\gamma}\approx 0.1$, but for low  $x_{\gamma}$ values the statistical errors are 
still important and in principle it is still possible to reach better
 precision at low $x_{\gamma}$.
\par
Figures \ref{fig:q-real} and \ref{fig:gluon} show that at least some
parametrisations of the photon pdf constrained by $e^+e^-$ measurements are
compatible with jet photoproduction in $ep$ as calculated in LO. We can consider 
this fact  as a demonstration of $e^+e^-$-$ep$ universality of the photon 
structure, but
taking into  account the  large errors involved, at the semi-quantitative level 
only.  
\subsection{Measurement of the effective parton density of the virtual photon}
\begin{figure}[htb]
\begin{center}
\psfig{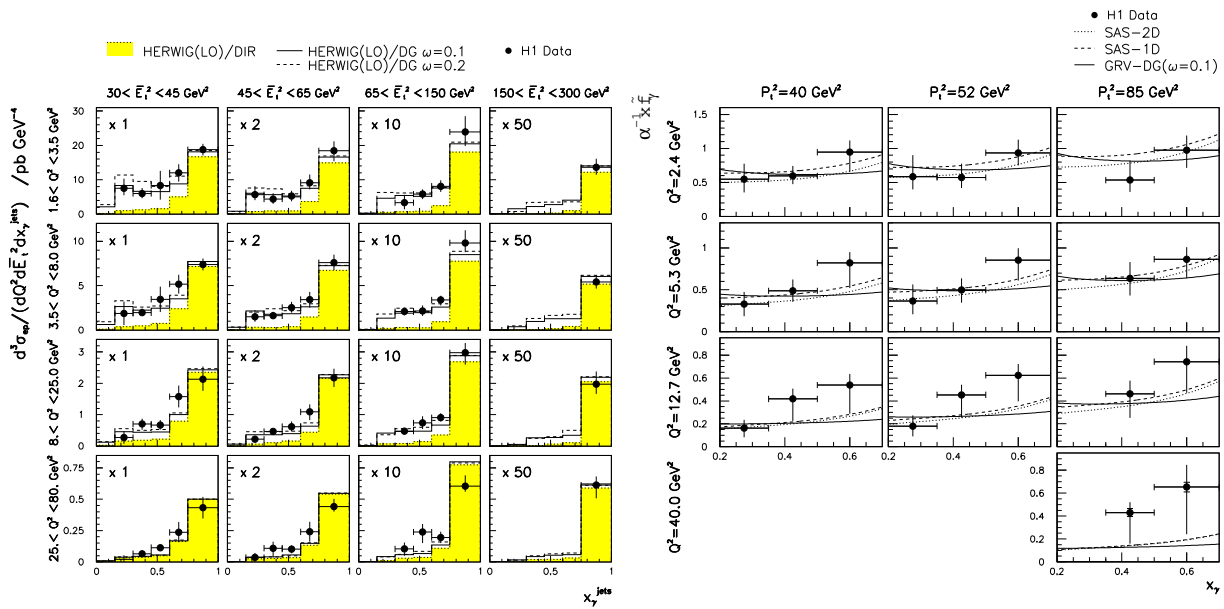}
\end{center}
\vspace*{-0.5cm}
\caption{{\em Left:} The differential di-jet cross section 
$d^3\sigma_{ep}/dQ^2d\bar{E}_t^2dx_{\gamma}^{jets}$ as a function of 
$x_{\gamma}^{jets}$ for different regions of $\bar{E}_t^2$ and $Q^2$.
The scale factors applied to cross sections are indicated. The error bar
shows the quadratic sum of systematic and statistical errors. The absence
of data point indicates that no measurement was made because of insufficient
statistics for the two dimensional unfolding. Also shown is the HERWIG(LO)/DG
model with $10\%$ soft-underlying event and two choices of the $Q^2$ 
suppression factor $\omega$. The full histogram is for $\omega=0.1$ ~{\rm GeV} and
dashed for $\omega= 0.2$ ~{\rm GeV}. The direct component is shown as a shaded 
histogram. {\em Right:} Leading order effective parton distribution function 
of virtual
photons extracted from triple differential cross section on the left.
 The data are compared to several theoretical predictions explained in the text.}  
\label{fig:triple}
\end{figure}

One can study the structure of virtual photons in a similar way to that  described in
 the previous section. Such a study \cite{H15} is presented in Figure 
\ref{fig:triple}  (left), based on sample of $6~{\rm pb}^{-1}$ of DIS events in the kinematic 
range
$1.6 \leq Q^2 \leq 80 ~{\rm GeV}^2$, $0.1 \leq y \leq 0.7$ with at least 2 jets with
average squared transverse energy $\bar{E}_t^2 > 30 ~{\rm GeV}^2$. Jets are restricted to 
be asymmetric in energy, and the constraints are such that neither jet has $E_T \leq 4 ~{\rm GeV}$
and the sum is always $\geq 11 ~{\rm GeV}$.

In Figure \ref{fig:triple} (left) the di-jet cross section is plotted as a function
of $x_{\gamma}^{jets}$ for different regions of $Q^2$ and $E_T^2$. One sees
a clear excess over the expectation of direct photon reactions, indicating that
virtual photons also have a resolved part. The direct interactions are manifest
in the region of $x_{\gamma}^{jets} \rightarrow 1$. With 
increasing $Q^2$ at fixed $E_T^2$ the direct contribution increases while
 the resolved is significant only for $E_T^2 > Q^2$ i.e. when the spatial extent of
the photon exceeds the resolution of the probe. This is a beautiful 
demonstration of QFT at work.
\par
It should be noted that pQCD does not provide a full theoretical description
of the virtuality  dependence of the parton density in the photon. 
There is only an asymptotic  
prediction \cite{walsh} in LO for $F_2^{{\gamma}^{\star}}$ i.e. the structure 
function of the virtual photon $\gamma^{\star}$
\begin{equation}
F_2^{\gamma} =3 \sum_{q}e_q^4\frac{\alpha}{\pi}x[x^2+(1-x)^2]{\rm ln}\frac{Q^2}
{P^2},
\label{eq:witten2}
\end{equation}
valid for $Q^2 \gg P^2 \geq \Lambda^2$.
Here $P^2$ is the photon virtuality while $Q^2$ is the virtuality of the probe
(resolution scale). The formula (\ref{eq:witten2}) is analogous to 
(\ref{eq:witten1})
for a quasi-real photon. Now, kinematics provides its own 
cut-off $P^2$ as long as
it exceeds $\Lambda^2$. The ``hadron piece'' in formula (\ref{eq:witten1})
no longer counts as $F_2^{VDM} \propto 1/P^4$.\footnote{Note that formulae
(\ref{eq:witten1}) and (\ref{eq:witten2}) refer to the photon structure function
as measured in $e^+e^-$ experiments, and there are conflicting notations:
in $e^+e^-$ $Q^2$ always denotes the virtuality of the probe (i.e. resolution
scale) while in $ep$ experiments at HERA $Q^2$ traditionally denotes the virtuality
of the photon which here happens to be a target. Throughout the article we stick
to traditional notations, leaving  the reader to determine the correct recognition of the
symbols.}
\par
For  the description of the triple differential di-jet cross section we can
employ phenomenological models of the   dependence of the 
effective parton density in the photon on its virtuality $Q^2$ \cite{drees}, 
\cite{schuler}, \cite{sjos}. The Drees-Godbole model (DG) \cite{drees} 
starts with  parton densities in the real photon and suppresses them by 
a factor ${\mathcal{L}}$
\begin{equation}
{\mathcal{L}}(Q^2,E_T^2,\omega^2)=\frac{{\rm ln}\frac{E_T^2+\omega^2}{Q^2+\omega^2}}
{{\rm ln}\frac{E_T^2+\omega^2}{\omega^2}},
\label{eq:DG} 
\end{equation}
where $\omega$ is a free parameter.
The quark densities in the real photon are suppressed by ${\mathcal{L}}$ and
the gluon densities by  ${\mathcal{L}}^2$. This ansatz,  based on analysis in
\cite{schuler}, is designed to interpolate smoothly between ${\rm ln}(Q^2)$ and
${\rm ln}(Q^2/P^2)$ in the LO asymptotic formulae for real 
(\ref{eq:witten1}) and virtual (\ref{eq:witten2}) photons. In
 the model of Schuler and Sj\"ostrand \cite{sjos}, the perturbative anomalous
component of the parton density in a virtual photon is designed to approach the 
asymptotic result (\ref{eq:witten2}) as in the DG model, while the VDM part is rapidly 
suppressed
by the factor $m_V^2/(m_V^2+Q^2)$. In this approach the  parton density 
in a quasi-real 
photon has to be initially decomposed into perturbative and non-perturbative
components, so that the description of the $Q^2$ suppression of the photon pdf
is inseparable from the parametrisation of the  quasi-real photon pdf. This is
not the case for the DG model, which can be applied to any quasi-real photon pdf.
\par
In Figure \ref{fig:triple}  (left) the predictions of the HERWIG MC with GRV 
photon pdf 
multiplied by the DG factor are compared with the data. Clearly the data are
well described by the  MC with suitable choice of the parameter $\omega = 0.2$
which controls the onset of the $Q^2$ suppression.
\par
In Figure \ref{fig:triple} (right) we show the effective parton density of 
the virtual
photon extracted from the triple differential di-jet cross section (on the 
left). In the whole measurement range the systematic 
errors dominate the total error. The $x_{\gamma}$ dependence of the
effective parton density shows a tendency to rise with increasing  $x_{\gamma}$
as predicted by formula (\ref{eq:witten2}). For quasi-real photons in a similar 
$x_{\gamma}$ range the pdf is flat. This fact can be interpreted as the effect of 
 the VDM contribution, which preferentially fills out the low  $x_{\gamma}$ region
flattening the otherwise rising  $x_{\gamma}$-dependence of the anomalous
pQCD contribution. With rising $Q^2$, the VDM contribution rapidly decreases, 
leaving only the rising  $x_{\gamma}$-dependence. The errors on the  
experimental data are clearly too large to differentiate between SAS and DG,
although only SAS has a rising tendency for reasons explained before (the DG model has 
no explicit VDM component).
\begin{figure}[hbt]
\begin{center}
\psfig{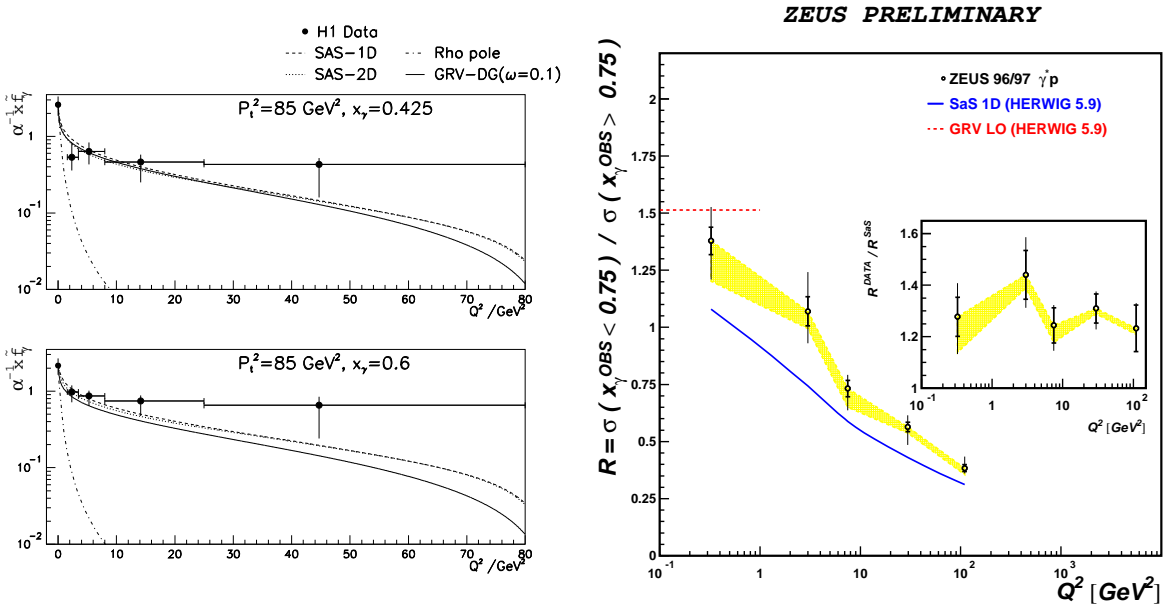}
\end{center}
\vspace*{-0.5cm}
\caption{{\em Left:} The leading order effective parton density of the photon 
as a function of $Q^2$ for $E_T^2=85 ~{\rm GeV}^2$ and two values of  $x_{\gamma}$.
Also shown are predictions from the DG model using GRV-LO pdf for a real photon
(solid line) and the SAS model (dotted line). The dot-dashed curve shows the
photoproduction data scaled by a $\rho$--pole factor. {\em Right:} The ratio of 
di-jet cross sections, $R^{DATA}=\sigma(x_{\gamma}^{OBS}
 < 0.75)/\sigma(x_{\gamma}^{OBS} > 0.75)$ as a function of photon virtuality
$Q^2$. The points represent ZEUS data. The predictions of HERWIG MC,
 $R^{SAS}=\sigma(x_{\gamma}^{OBS}< 0.75)/\sigma(x_{\gamma}^{OBS}> 0.75)$
using the SAS 1D photon pdfs, are shown as a full line. The ratio $R^{DATA}/
R^{SAS}$ as a function of $Q^2$ is also shown in the inset.}
\label{fig:q2dep}
\end{figure}
In Figure \ref{fig:q2dep} (left) we show the extracted photon pdf as a function of
$Q^2$. The data are compared with the pure VDM prediction and the DG and SAS models
discussed above. The DG and SAS parametrisations describe the data up to
$Q^2 \approx 25 ~{\rm GeV}^2$. The last data point $25 \leq Q^2 \leq 80 ~{\rm GeV}^2$
is above both parametrisations. This is however the region where $Q^2 \rightarrow
E_T^2$ and non-leading terms not accounted for in the models are expected to be
important and may affect the extraction of the effective parton distribution
from the data (breakdown of factorisation).
\par
The ZEUS collaboration investigated the photon pdf suppression with increasing $Q^2$
in a different way. They use an operational definition of direct and resolved 
di-jet  cross sections:
\begin{math}
\sigma^{dijet}({\rm direct})=\sigma^{dijet}(x_{\gamma}^{jet} > 0.75),
\end{math}
\begin{math}
\sigma^{dijet}({\rm resolved})=\sigma^{dijet}(x_{\gamma}^{jet} < 0.75).
\end{math}
 In Figure \ref{fig:q2dep} (right) we show the ratio of the so-defined 
``resolved'' to 
``direct'' contributions as a function of $Q^2$. The SAS prediction is 
below the data points, but the shape of the dependence is correct.
\par
At the end of this section we would like to underline that there is no basic
difficulty in the theoretical description of the region $E_T^2 \approx Q^2$. On
 the contrary, in this region of phase space, terms with ${\rm ln}(E_T/Q)$ appearing
in the perturbative expansion are small, so that even fixed order theory is
applicable in this case. It is only the photon structure function approach
which breaks down for  $Q \approx E_T$. Let us also note that in the
range $E_T^2 \gg Q^2 \gg \Lambda^2$ the photon structure approach can,
in principle, be replaced by resummed $({\rm ln}(E_T^2)/Q^2))^n$ theory, because for
 the photon, unlike for hadrons, the purely perturbative 
(anomalous) part of pdf
dominates completely at high values of $E_T^2$. In this sense the structure
function of the virtual photon could and should be considered as a
convenient parametrisation of higher order QCD corrections which are not 
available yet.
\begin{figure}[p]
\vspace{-1cm}
\begin{center}
\centerline{\psfig{file=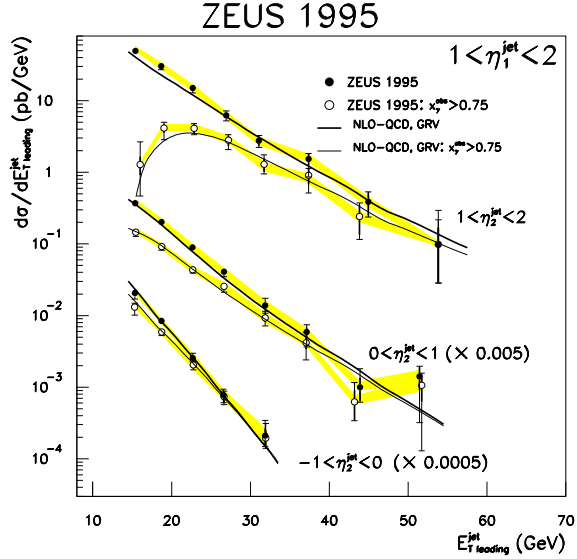,height=7.5cm}}
\end{center}
\caption{Di-jet cross section as a function of $E_{Tleading}^{jet}$ for
 $\eta_1^{jet}$ between 1 and 2 and several bins of  $\eta_2^{jet}$. The
filled circles correspond to the entire range while open circles correspond
to ``direct'' events with ${\gamma}^{jet} > 0.75$. The data are compared to
NLO-QCD calculations, using the GRV-HO parametrisation for the photon 
structure.}
\label{fig:zeusdijet}
\begin{center}
\centerline{\psfig{file=zeusjet.epsi,height=8cm}}
\end{center}
\caption{{\em Left:} Differential cross sections in $x_{\gamma}^{obs}$ in 
slices of the transverse energy of the leading jet $E_T^{jet1}$. The data
points are shown with statistical errors (inner bars) and statistical and
systematic errors added in quadrature (outer bars). The scale uncertainty
is shown as the shaded band. The data are compared to a NLO prediction which
uses the AFG-HO photon structure function. {\em Right:} The relative
difference in cross sections between data and NLO predictions for 
distributions in the Figure on the left. The estimation of the scale uncertainty
in the NLO calculation, obtained by varying the factorisation and 
renormalisation scales by a factor two, is shown as the diagonally shaded 
band.}
\label{fig:zeusjet}
\end{figure}

\section{Comparison of the measured di-jet cross section with a NLO QCD 
calculation}
The ZEUS Collaboration analysis of di-jet photoproduction 
\cite{Z3}, \cite{Z2},  based on a large sample of events, confronts
a precision measurement with a NLO QCD calculation. The cut on the transverse energy of jets 
$E_T^{jet1,2} > 14,11 ~{\rm GeV}$ is so high that corrections for hadronisation and the 
soft underlying event are certainly small (less than $10 \%$). Therefore
the data, corrected only for  detector effects, are compared directly to NLO
QCD calculations \cite{kramer} at the parton level. There is good general
agreement of the di-jet cross section with the NLO calculation as seen e.g. in
Figure \ref{fig:zeusdijet} where it is plotted as a function of the 
transverse energy. While $E_{Tleading}^{jet}$ increases from 14 to 50 ~{\rm GeV}, cross section falls by 3 orders of magnitude and no obvious deviations
from theory are seen, except for events with forward jets 
$1 < \eta_{1,2}^{jet} < 2$
and $E_{Tleading}^{jet} < 25 ~{\rm GeV}$. Let us note that the 
theory--data discrepancy
appears exactly in the region where the distance between ``direct'' and ``direct +
resolved'' points is large, meaning that the influence of the resolved 
component
is large. This tendency of the data becomes even more evident when the di-jet
cross section is plotted as a function of the photon momentum fraction 
\cite{Z3}.
In Figure \ref{fig:zeusjet} (left)  the di-jet cross section is plotted as a 
function
of $x_{\gamma}^{obs} \equiv x_{\gamma}^{jet}$  in several bins of the
transverse energy of the leading jet. Except for the last bin (dominated by
direct process) all data points are above the NLO cross section.
\par
Figure \ref{fig:zeusjet} (right) shows the relative value of the theory-data
discrepancy  as a function of  $x_{\gamma}^{obs}$. The relative discrepancy
reaches 60$\%$. Even taking into account the uncertainty of the theory due to
 the choice of the NLO scale (shown as shaded area) and the 
experimental uncertainty 
due to the hadronic energy scale, we have to admit that there is an indication 
that the parton densities in the photon are underestimated. Incidentally, this 
conclusion seems to be confirmed by recent measurements of the di-jet cross 
section in $\gamma^{\star}\gamma^{\star}$ collisions \cite{OPAL} . Even so 
it is difficult to accept without reservation that the pdf in photon is
much underestimated. The argument against such interpetation of ZEUS and OPAL
results is the  general good agreement of the LEP  measurements of photon 
structure function with GRV parametrisation. This
problem certainly deserves further scrutiny.  

\section{Summary and Conclusion}
 Measurements of $e\gamma$ DIS and jet electroproduction at HERA confirm the 
simple picture of photon structure first identified in $e^+e^-$ experiments
at PETRA. There is general  compatibility of $e^+e^-$ and $ep$ measurements
which demonstrates the validity of a factorisable structure which can be 
assigned to the photon, both real and virtual. Properties of the photon structure
function are in agreement with theoretical expectations, in particular
its dependence on the photon momentum fraction and resolution scale. Also the  
parton density suppression with increasing photon virtuality is in agreement
with phenomenological models based on asymptotic QCD predictions.
\par
 The recent precision measurements of the di-jet cross section seem to 
indicate that 
the existing parametrisations (based on $e^+e^-$ measurements) of the photon 
pdf are inadequate for a precise description of the data using NLO theory.
This problem certainly deserves further scrutiny.  
\par
 The accumulated  LEP, LEP2 and HERA data with much reduced systematic 
uncertainties over a very wide kinematic range now wait for a combined 
analysis, 
which would result in a new NLO parametrisation of the photon pdf. The concepts
and tools for such an analysis have yet to be developed.   
 
%


\begin{thebibliography}{99}
\bibitem{dainton} J. Dainton, Invited presentation at the discussion 
meeting {\em The Quark Structure of Matter}, Royal Society, London, 
May 24 2000, to appear in Philosophical Transactions of the Royal Society 
of London Series A: Mathematical, Physical and Engineering Sciences,
[hep-ex/0009033].
\vspace{-2.5mm}
\bibitem{krawczyk} M. Krawczyk, A. Zembrzuski and M. Staszel, {\em Survey of Present Data on 
Photon Structure Functions and Resolved Processes} to appear in 
{\it Phys. Rep.},   [hep-ph/0011083].
\vspace{-2.5mm}
\bibitem{nissius} R. Nisius, 
{\em The photon structure from deep inelastic electron-photon scattering},
 [hep-ex/9912049].
\vspace{-2.5mm}
\bibitem{alevi} A. Levy, {\em The proton and the photon, who is probing whom in electroproduction?} [hep-ph/0002015].
\vspace{-2.5mm}
\bibitem{OPAL2} G. Abbiendi et al. (OPAL Collaboration), \epj{C18}{2000}{15}.
\vspace{-2.5mm}
\bibitem{melles}M. Melles, \np{B (Proc. Suppl.) 82}{2000}{379}.
\vspace{-2.5mm}
\bibitem{ioffe} B.L. Ioffe, \pl{B30}{1969}{123}; ~B.L. Ioffe, V.A. Khoze
and L.N. Lipatov, {\em Hard Processes}, (North-Holland, 1984).
\vspace{-2.5mm}
\bibitem{VDM} J.J. Sakurai, \prl{22}{1969}{981}.
\vspace{-2.5mm}
\bibitem{witten} E. Witten, \np{B 120}{1977}{189}.
\vspace{-2.5mm}
\bibitem{altar} G. Altarelli, G. Parisi, \np{B126}{1977}{298}.
\vspace{-2.5mm}
\bibitem{LAC} H. Abramowicz, K. Charchula and A. Levy, \pl{B269}{1991}{458}.
\vspace{-2.5mm}
\bibitem{GRV} M. Gluck, E. Reya and A. Vogt, \prev{D46}{1992}{1973}.
\vspace{-2.5mm}
\bibitem{AGF} P. Aurenche, J.P. Guillet and  M. Fontannaz, \zp{C64}{1994}{621}.
\vspace{-2.5mm}
\bibitem{SAS} G.A. Schuler and  T. Sj\"ostrand,  \np{B 407}{1993}{539}.
\vspace{-2.5mm}
\bibitem{OPAL} T. Wengler, {\it The hadronic picture of the photon}, Proceedings
of 30th International Conference on High Energy Physics (ICHEP 2000),Osaka.
\vspace{-2.5mm}
\bibitem{H11} T. Ahmed et al (H1 Collab.), \np{B445}{1995}{195}.
\vspace{-2.5mm}
\bibitem{H12} C. Adloff et al (H1 Collab.), \pl{B415}{1997}{418}.
\vspace{-2.5mm}
\bibitem{H13} C. Adloff et al (H1 Collab.), \epj{C1}{1999}{97}.
\vspace{-2.5mm}
\bibitem{H14} C. Adloff et al (H1 Collab.), \epj{C10}{1999}{363}.
\vspace{-2.5mm}
\bibitem{H15} C. Adloff et al (H1 Collab.), \epj{C13}{2000}{397}.
\vspace{-2.5mm}
\bibitem{Z1} J. Breitweg et al (ZEUS Collab.), \epj{C11}{1999}{35}.
\vspace{-2.5mm}
\bibitem{Z2} J. Breitweg et al (ZEUS Collab.), \epj{C6}{1999}{67}.
\vspace{-2.5mm}
\bibitem{Z3} ZEUS Collab., {\it The Structure of the Photon and the Dynamics of  Resolved Photon 
Processes in Di-jet Photoproduction at HERA}, Proceedings
of 30th International Conference on High Energy Physics (ICHEP 2000),Osaka. 
\vspace{-2.5mm}
\bibitem{kramer}M. Klasen, T. Kleinwort and  G. Kramer, \epj{C1}{1998}{1}.
\vspace{-2.5mm}
\bibitem{comb}B.V. Combridge and  C.J. Maxwell, \np{B239}{1984}{429}.
\vspace{-2.5mm}
\bibitem{walsh}T. Uematsu and  T.F. Walsh, \np{B199}{1982}{93}.
\vspace{-2.5mm}
\bibitem{drees}M. Drees and  R. Godbole, \prev{D50}{1994}{3124}.
\vspace{-2.5mm}
\bibitem{sjos}T. Sj\"ostrand and  G.A.Schuler, \pl{B376}{1996}{193}.
\vspace{-2.5mm}
\bibitem{schuler}F. Borzumati and  G.A. Schuler, \zp{C58}{1993}{139}.
\vspace{-2.5mm}
\end{thebibliography}
\end{document}